\documentclass[aps, pre, notitlepage, superscriptaddress, twocolumn,
               a4paper, floatfix]{revtex4-1}

\usepackage[utf8]{inputenc}
\usepackage{epsfig}
\usepackage[T1]{fontenc}
\usepackage{t1enc}
\usepackage{graphicx}
\usepackage{amssymb}
\usepackage{amsmath}
\usepackage{relsize}
\usepackage[caption=false]{subfig}
\usepackage{color}

\usepackage[normalem]{ulem}

\usepackage{mathtools}
\def\multiset#1#2{\ensuremath{\left(\kern-.3em\left(\genfrac{}{}{0pt}{}{#1}{#2}\right)\kern-.3em\right)}}

\global\long\def\vek#1{\boldsymbol{#1}}

\begin{document}

\title{ Aggregation and fragmentation dynamics in random flows: From
  tracers to inertial aggregates.}

\author{Ksenia Guseva}
\email{ksenia.guseva@uni-oldenburg.de}
\affiliation{Theoretical Physics/Complex Systems, ICBM, University of Oldenburg, 26129 Oldenburg, Germany}

\author{Ulrike Feudel}
\email{ulrike.feudel@uni-oldenburg.de}
\affiliation{Theoretical Physics/Complex Systems, ICBM, University of Oldenburg, 26129 Oldenburg, Germany}

\pacs{}

\begin{abstract}
  We investigate aggregation and fragmentation dynamics of tracers and inertial
  aggregates in random flows leading to steady state size distributions. Our
  objective is to elucidate the impact of changes in aggregation rates, due to
  differences in advection dynamics, especially with respect to the influence of
  inertial effects. This aggregation process is, at the same time, balanced by
  fragmentation triggered by local hydrodynamic stress. Our study employs an
  individual-particle-based model, tracking position, velocity and size of each
  aggregate. We compare the steady-state size distribution formed by tracers and
  inertial aggregates, characterized by different sizes and densities. On the
  one hand, we show that the size distributions change their shape with changes
  of the dilution rate of the suspension. On the other hand, we obtain that the
  size distributions formed with different binding strengths between monomers
  can be rescaled to a single form with the use of a characteristic size for
  both dense inertial particles and tracer monomers. Nevertheless, this last
  scaling relation also fails if the size distribution contains aggregates that
  behave as tracer-like and as inertial-like, which results in a crossover
  between different scalings.
  
\end{abstract}

\maketitle

\section{Introduction}

Aggregation and fragmentation dynamics in turbulent flows is found at the core
of many processes in environmental and engineering sciences. Examples of such
systems include the formation of rain droplets in
clouds~\cite{falkovich_acceleration_2002}, the settling of marine aggregates in
estuaries and open ocean~\cite{burd_particle_2009, maerz_modeling_2011}, and
even growth of stars and galaxy formation by dust particle
collisions~\cite{silk_star_1980}. In all these cases, the balance between
aggregation and fragmentation dynamics determines the characteristics of the
population of particles --- such as aggregate size distribution --- in the
steady-state. While different mechanisms can be identified to be responsible for
breakups (e.g. instability after reaching a certain size, or due to the action
of hydrodynamic forces), aggregation always results from collisions, and hence
depends on the advection of aggregates by the velocity field.

In this work we analyze the steady-state size distribution produced by the
competition between aggregation and breakup by hydrodynamic stress for
aggregates subjected to different advection dynamics (tracers in contrast to
inertial aggregates). We use an {\it individual-particle-based} model where each
aggregate is moved independently in space, and at each instant in time is
identified with a position, a velocity and a size. This approach was previously
used to investigate the influence of different fragmentation mechanisms on
aggregation-fragmentation dynamics of small inertial aggregates in
random~\cite{zahnow_what_2009, zahnow_particle-based_2011} and convection flows
~\cite{zahnow_coagulation_2009, zahnow_aggregation_2008}.  Here we choose a
single mechanism for fragmentation --- breakup due to hydrodynamic stress, and
focus on the influence of changes in aggregation rates on the steady-state size
distribution. With this aim, we consider particles that move differently while
advected by the flow field and therefore have distinct collision (aggregation)
rates. These types of particles can be classified into tracers and inertial
aggregates. Tracers are particles that follow the velocity field exactly,
identically to fluid elements. Inertial aggregates, on the contrary, deviate
from the fluid trajectories due to action of several hydrodynamic forces.  One
important characteristics unique to inertial aggregates is that their motion
depends strongly on their size.

Several recent studies have theoretically evaluated the aggregation rates for
both tracers~\cite{saffman_collision_1956} and for inertial
particles~\cite{bec_clustering_2005} in the absence of fragmentation
dynamics. For inertial particles in particular, the collision dynamics is known
to be strongly enhanced in turbulence due to phenomena such as {\it preferential
  concentration}~\cite{eaton_preferential_1994, wang_settling_1993} and {\it
  caustics}~\cite{falkovich_acceleration_2002, wilkinson_caustics_2005,
  wilkinson_path_2003}, for a review see~\cite{pumir_collisional_2016}. Other
studies have estimated fragmentation rates, when these are triggered by
hydrodynamic forces in the absence of aggregation, although only for tracer
particles~\cite{babler_breakup_2012, babler_numerical_2015}. Aggregation and
fragmentation processes are brought together only within a mean field
approximation using {\it population balance equations}, which are able to
determine the size distribution as a function of time for a reversible
aggregation process~\cite{spicer_coagulation_1996, babler_breakup_2012,
  babler_analysis_2007, ben-naim_phase_2008}. Although significant progress can
be achieved with this approach, it cannot account for spatial fluctuations in
particle numbers, which are enhanced by inertial effects and for the
location-dependent fragmentation rates.  Therefore we propose to treat the
aggregation and fragmentation as coupled processes, taking fully into account
the spatial variations in distribution and sizes of aggregates by means of an
{\it individual-particle-based} model.

Our aim is to investigate the consequences of inertia on the steady-state of
aggregation and fragmentation processes. We are interested in the case where
breakup of aggregates occurs due to hydrodynamic forces, using a random flow
field mimicking turbulence. We use different ensembles of inertial monomers
(primary, unbreakable particles), and explore the influence of Stokes number and
density on the steady state, which result from differences in advection
dynamics.  We show the changes induced in the dynamics and their results in the
scaling of the mean size and other characteristics of the size distribution.  We
analyze the scaling properties of the size distribution with respect to the
resistance of aggregates to hydrodynamic forces and to the dilution of the
suspension for ensembles of aggregates with different inertial properties.

This paper is organized as follows. Sec.~\ref{sec:model_all} consists of
implementation details of the {\it individual-particle-based} model. It is
divided in Subsect.~\ref{sec:MR}, where we describe the advection of inertial
particles, and Subsects.~\ref{sec:agg-frag} and \ref{sec:shear}, where we
describe the details for the aggregation and fragmentation processes. We proceed
with the results for the influence of the dilution rate of the suspension for
the aggregation-fragmentation dynamics of tracer and of inertial aggregates in
Sec.~\ref{sec:dilution_rate}.  Finally we analyze the effect of the binding
strength between monomers in Subsect.~\ref{sec:binding_tracers} and
Subsect.~\ref{sec:binding_inertial}, again comparing the steady state properties
of ensembles of aggregates with different advection dynamics. The comparison and
a summary of the impact of advection dynamics and results for other ensemble
types are presented in Subsect.~\ref{sec:tr_vs_inertial}. We conclude and then
discuss the results in Sec.~\ref{sec:conclusion}.

\section{Methods and theoretical introduction}\label{sec:model_all}

We start by presenting details of the three dynamical processes which compose
our model: advection, aggregation and fragmentation. Our system consists of
particles which are advected by the flow while aggregating with each other upon
collisions, and fragmenting due to local forces in the flow field. This section
is divided into three parts: first we describe the advection dynamics in
Subsect.~\ref{sec:MR}; then we proceed in Subsect.~\ref{sec:agg-frag} with the
general aspects of the aggregation and fragmentation dynamics; finally in
Subsect.~\ref{sec:shear} we provide details of the forces that drive
fragmentation events.

\subsection{Advection of inertial particles}\label{sec:MR}

We assume that all aggregates are immersed in a moving fluid and advected by
it. As the flow field, $\mathbf{u} (\mathbf{X}, t)$, we use a random (synthetic)
flow~\cite{sigurgeirsson_algorithms_2001, garcia-ojalvo_generation_1992,
  marti_langevin_1997} which mimics homogeneous isotropic turbulence in the
dissipative scale. It corresponds to a Gaussian incompressible velocity field
generated by means of random perturbations of Fourier modes. Details of the
implementation of this random flow field can be found in Appendix I. The
important parameters which characterize the flow are: $\tau_f$ -- correlation
time; $\lambda_f$ -- correlation length; and $u_0$ -- mean of the absolute
velocity. These three parameters can be combined into a single dimensionless
parameter: the Kubo number, $Ku = u_0\tau_f/\lambda_f$, which characterizes the
persistence of coherent structures in the flow field with respect to the time
needed for fluid particles to explore them~\cite{pumir_collisional_2016}.

\begin{figure}[h]
  \includegraphics[scale=.75]{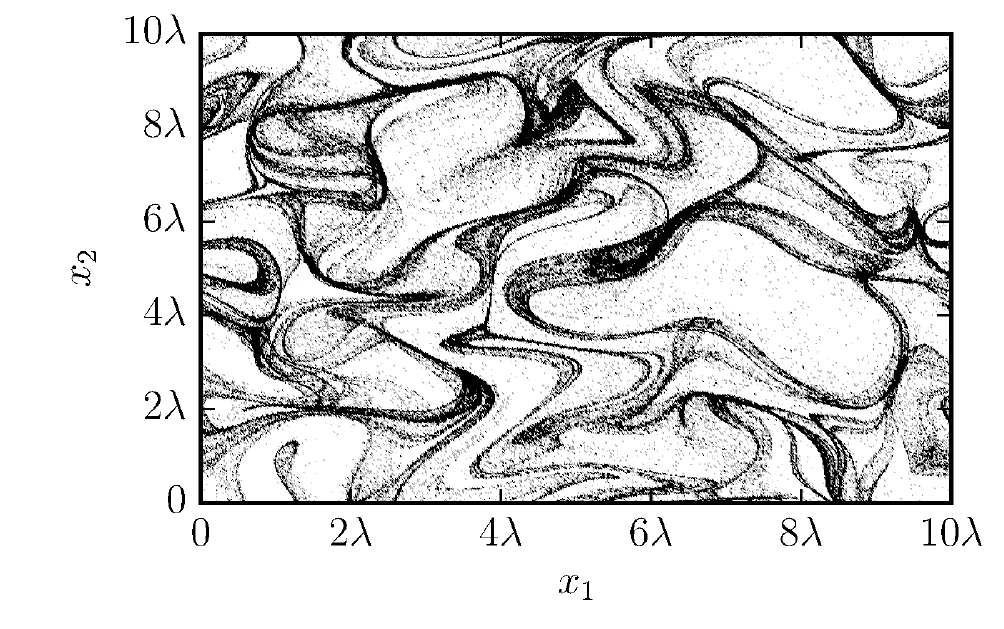}
  \caption{Non-homogeneous spatial distribution of inertial particles, with
    $St = 0.5$ for a flow field with $Ku = 1$.}\label{fig:pref_conc}
\end{figure}

We assume a dilute suspension of small spherical particles, which are much
smaller than the smallest coherent structures (eddies) of the flow. On the one
hand we consider particles which follow the flow field exactly.  Their
trajectories can be obtained from: $\dot{\vek X} ={\vek u(\vek{X}, t)}$. These
particles are called {\it tracers}. On the other hand, we look at particles
whose trajectories may deviate from flow field trajectories due to inertial
effects. The motion of these inertial particles can be described by the
following equations
\begin{equation}\label{eq:MR}
  \dot{\vek X} ={\vek V} \qquad \qquad \dot{\vek V}=\beta D_t \vek u(\vek{X}, t)-\frac{1}{\tau_p}\left(\vek v-\vek u(\vek{X}, t)\right),
\end{equation}

where $\mathbf{X}$, $\mathbf{V}$ correspond to the position and the velocity of
a particle respectively. Eq.(\ref{eq:MR}) are an approximate form of the
Maxey-Riley equations and are valid in the limit of small particle Reynolds
numbers. They were formulated independently by
Gatinol~\cite{gatignol_faxen_1983}, Maxey \& Riley~\cite{maxey_equation_1983}
and later corrected by Auton~\cite{auton_force_1988}.  The abbreviation
$D_t{ } = \partial_t + \mathbf{u}\cdot\nabla$ represents the derivative along
the flow trajectory. The ratio between aggregate's $\rho_0$ and fluid's $\rho_f$
densities is given by the parameter $\beta = \frac{3\rho_f}{\rho_f
  +2\rho_0}$. The time it takes for an aggregate to adjust to the changes of the
flow is given by the Stokes time $\tau_{p}$, which depends on the aggregate's
size $r_p$. This response time can be compared to the smallest time scale of the
flow $\tau_f$. The relation between these two time scales gives a dimensionless
number, the Stokes number:
\begin{equation}
  St = \frac{\tau_{p}}{\tau_f} = \frac{r_{p}^2}{3\beta \nu}\frac{1}{\tau_f},
\end{equation}
with $\nu$ as the kinematic viscosity. While tracers are distributed
homogeneously in space, inertial particles segregate into random
attractors due to dissipation (Fig.~\ref{fig:pref_conc}). For a review
of the dynamics of inertial particles in random flows, and the
mechanisms which drive the segregation such as {\it preferential
  concentration}~\cite{eaton_preferential_1994,
  wang_settling_1993,bec_acceleration_2006, sigurgeirsson_model_2002}
and random amplification~\cite{gustavsson_ergodic_2011}
see~\cite{pumir_collisional_2016}.

The segregation into fractal attracting regions is a well established
explanation for the fast growth of rain droplets in clouds, since it would
strongly increase the number of collisions and consequently the aggregation
rates~\cite{falkovich_acceleration_2002, pinsky_turbulence_1997}. This increase
has already been demonstrated for inertial particles in random flows
~\cite{bec_clustering_2005}. Our objective is to explore the consequence of an
increased aggregation rate on size distributions which results from aggregation
and fragmentation dynamics.

\subsection{Aggregation-fragmentation dynamics}\label{sec:agg-frag}

The dynamics of each aggregate is modeled {\it individually}, and we attribute
to each one of them, at each instant of time, --- a position, a velocity and a
size. For convenience the sizes of all aggregates in the system are measured
with respect to a unit particle --- the monomer. We assume that all monomers are
small spheres with a radius $r_0$ and mass $m_0$
($m_0 = \frac{3\rho_0}{4\pi r_0^3}$, where $\rho_0$ is the monomer's
density). Their main property is that they cannot be broken. However they can
form large aggregates since they can stick together upon collision and hence,
build an ensemble of aggregates with different sizes in the
suspension. Therefore, each aggregate in the system can be described by an {\it
  integer} number $\alpha$ of monomers that compose it. In other words, $\alpha$
establishes a relation between the size of an aggregate and the unit size, hence
specifying the aggregate's mass as $m_{\alpha} = \alpha m_0$, and its radius as
$r_{\alpha} = \alpha^{1/3} r_0$. Here we assume that each aggregate is a sphere.
According to this each of the aggregates has its specific Stokes number
$St_{\alpha} = \frac{r_{\alpha}^2}{3\beta \nu}\frac{1}{\tau_f},$

While two aggregates, of radius $r_i$ and $r_j$ (with $i$ and $j$ monomers
respectively) move, they may approach each other. When the relative distance $d$
between them becomes equal to $r_i + r_j$ a collision occurs. We assume that
each of such collision events, results in aggregation, where a new particle of
size $\alpha = i + j$ is formed. We postulate that in all aggregation events the
mass and the momentum are conserved, and from this condition we can derive all
the properties of the resulting aggregate. Furthermore, all aggregates have a
spherical shape independently of their size and the same density, $\rho_0$, as
the monomers which form them. To be able to track all the collisions efficiently
we use an event-driven algorithm, a detailed description of it can be found
in~\cite{sigurgeirsson_algorithms_2001}.

Fragmentation, alternatively, is driven by an external force exerted on the
aggregate, which depends on the position of the aggregate in space; the details
about this force are presented in Subsec.~\ref{sec:shear}. In these breakup
events two smaller aggregates of similar size are produced (binary
fragmentation), so that mass conservation is obeyed. These fragments are placed
withing a close distance of each other. Their velocity is kept the same as of
the aggregate before break-up to ensure momentum conservation. Fragmentation and
aggregation are competing processes, and the balance between them leads to a
steady state size distribution of aggregates in the suspension, containing
$N_{\alpha}$ aggregates with a particular size $\alpha$.

\begin{figure*}[t!]
  \includegraphics[scale=.73]{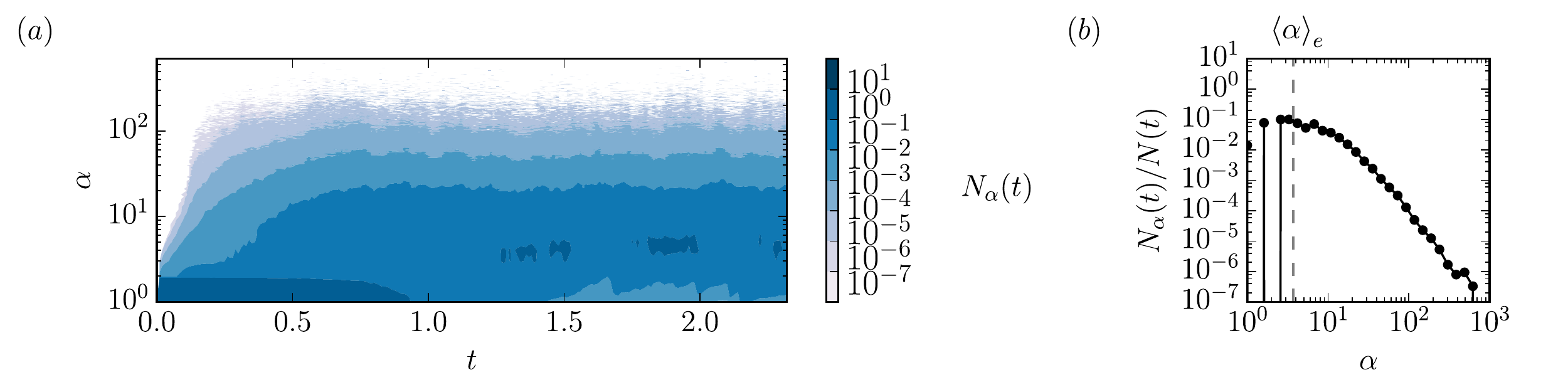}
  \caption{(a) Evolution of the size distribution in time. (b) size distribution
    of aggregates in the steady state, the dashed gray line represents the mean
    aggregate size $\left<\alpha\right>_e$. $N$ is the total number of
    aggregates in the system
    ($N = \sum_{\alpha} N_{\alpha}$).}\label{fig:ex_sizedistrib}
\end{figure*}

We initialize our suspensions always with homogeneously spread monomers,
$N_1(0) = M$ and $N_{\alpha} = 0$ with $\alpha \neq 1$, where $M$ is the total
number of monomers. Then we track the time evolution of the number of aggregates
in each size class $N_{\alpha}(t)$ resulting from advection of monomers and
aggregation, see Fig.~\ref{fig:ex_sizedistrib} (a). The transient dynamics
consists of an initial period of irreversible aggregation, followed by an
increase in fragmentation events. These breakage events become more and more
relevant as the sizes of aggregates in the suspension increase. The number of
fragmentation events grows until, on average, it balances all aggregation events
and the dynamics reaches a steady state. The time to reach this balance depends
again on the number and properties of monomers initialized in the
suspension. The resulting steady state size distribution is characterized by an
average aggregate size $\left<\alpha\right>_e$ and a size distribution of
aggregates, Fig.~\ref{fig:ex_sizedistrib} (b). To eliminate time fluctuations of
$\left<\alpha\right>_e$, we define a time average computed for the interval $T$
(after discarding transients), i.e.
$\left<\; \right>_t = \int_{t}^{t+T} \left< \right>_e(t') dt'$. In
Fig.~\ref{fig:ex_sizedistrib} (a, b) we show an example of the evolution of the
size distribution in time and the corresponding steady state size distribution
respectively.

It is interesting to note that the traditional mean field approach for
aggregation-fragmentation dynamics using certain kernels predicts that the size
distribution in the steady state obeys a scaling relation with respect to a
characteristic particle size~\cite{family_kinetics_1986, ernst_scaling_1987,
  meakin_scaling_1988, meakin_aggregation_1992}. In our case the prediction is
\begin{equation}
N_{\alpha} = \frac{1}{\alpha^2}f(\alpha/\left<\alpha\right>_e).\label{eq:scaling}
\end{equation} 

Additional scaling properties can also be derived for $\left<\alpha\right>_e$,
establishing that it scales as a power law with the total number of monomers in
the suspension $M$. Moreover, it scales as a power law with the strength of the
binding forces inside the aggregate. Our objective is to verify if this scaling
form of the size distribution holds also for our individual particle approach,
which takes the spatial distribution of aggregates explicitly into account. By
contrast, the mean filed approach is based on aggregation rates which are
independent of the spatial distribution of aggregates.

\subsection{Fragmentation by hydrodynamic stress}\label{sec:shear}

\begin{figure}[h]
  \includegraphics[scale=.95]{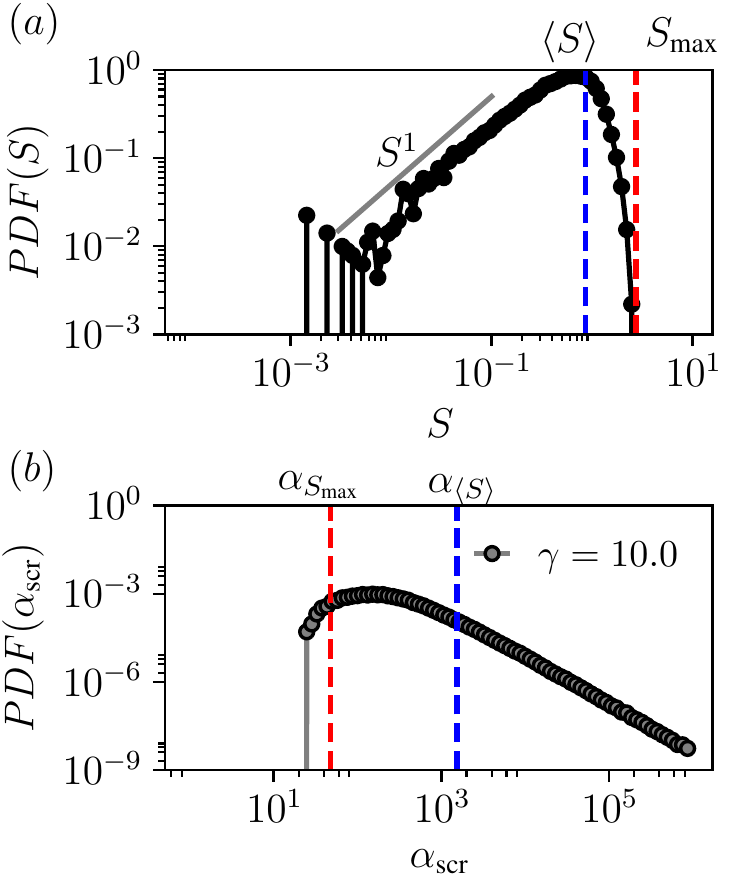}
  \caption{(a) PDF of the shear forces for a random flow, obtained from space
    sampling of $\mathbf{u} (\mathbf{X},t)$, with $Ku = 1$ ($\tau_f = 1$) and
    (b) the corresponding maximum aggregate's size allowed by these forces, the
    parameters used are $\gamma = 10$ and $\kappa = 3$.}\label{fig:hist_shear}
\end{figure}

We are interested in a specific type of fragmentation process in which breakup
events develop from interactions of aggregates with the local flow field. In our
study the flow is responsible for two types of dynamics: (i) it carries
particles around so that they can meet and aggregate and; (ii) it exerts forces
that act on the particle's structure, being able to trigger breakups. For such
type of fragmentation, numerous studies point to a strong link between
fragmentation rates of aggregates and their morphological properties, such as
shape, porosity and more importantly their size, see~\cite{zaccone_breakup_2009,
  de_bona_internal_2014}. The hydrodynamic stress that drives the breakup events
is the shear force, defined as
$S = (2\sum_{i}\sum_{j}S_{ij}S_{ij})^{\frac{1}{2}}$, where
$S_{ij} = \frac{1}{2}\left(\frac{\partial u_i}{\partial x_j} + \frac{\partial
    u_j}{\partial x_i}\right)$ corresponds to the rate-of-strain tensor of the
flow field, $\mathbf{u} (\mathbf{X},t)$~\footnote{It is important to note that
  the fragmentation by hydrodynamic forces is observed to be the most relevant
  fragmentation mechanism only for light particles}. The shear force varies in
space and time, and has predefined statistical properties, such as a probability
distribution~\footnote{In particular, for a turbulent flow, it is also useful to
  relate the shear, to the dissipation rate by $\epsilon = \nu S$, where $\nu$
  corresponds to the kinematic viscosity of the fluid}.  For the random flow
used in this study, this probability distribution of the shear forces is shown
in Fig.~\ref{fig:hist_shear} (a).

As described previously we prescribe the existence of a smallest aggregate's
size, the monomer, which cannot be broken by the shear forces of the flow
field. All other, larger aggregates in the suspension produced by means of
aggregation of these monomers are subject of fragmentation. The monomers inside
an aggregate are hold together by internal binding forces. When the shear forces
of the flow field, which act on the aggregate, are stronger than these binding
forces a fragmentation event occurs. Therefore we can define a critical shear
$S_{\text{cr}}$ at which the shear force becomes larger than the binding forces
resulting in fragmentation. Importantly, this $S_{\text{cr}}$ depends on the
size of the aggregate. Moreover, the functional form of the critical shear
relies on the assumption that large aggregates are more likely to break than the
smaller ones and therefore are more sensitive to fluctuations of the shear in
the flow.

We use a power law relation between the critical shear and the aggregate size,
\begin{equation}
S'_{cr} = \gamma' \alpha^{-\frac{1}{\kappa}},\label{eq:s_cr}
\end{equation}
where $\alpha$ is the number of monomers in the aggregate, and the additional
parameter $\gamma'$ and the exponent $\kappa$ represent the dependence on other
morphological properties. This relation, taken from
literature~\cite{parker_floc_1972, jarvis_review_2005}, results from numerical
simulations of individual aggregates as well as experiments, by taking into
account the complex fractal structure of aggregates and its internal binding
forces, in different flow fields. We assume that all aggregates fragment into
two smaller ones with similar sizes (binary fragmentation). The time scale on
which a breakup occurs is supposed to be significantly shorter than all other
relevant time scales of the dynamics. Therefore we approximate each breakup
event as instantaneous, and assume that it occurs exactly at the point where the
aggregate experiences a shear stress that exceeds the critical
value. Furthermore, if the shear stress is still larger than the binding forces
of any of produced fragments, another fragmentation event is performed within
the same time step. This can result in a fragmentation cascade, which proceeds
until all produced fragments are able to resist to local shear. For convenience
we work with a relative critical shear value, by comparing it with the mean
shear rate of the flow $\left<S\right>$. We assign
$S_{cr} = S_{cr}'/\left<S\right> = \gamma \alpha^{-\frac{1}{\kappa}}$, with
$\gamma = \gamma'/ \left<S\right>$. Note that, with the use of this rescaling,
$\gamma \leq 1$ corresponds to the case where for any $\alpha$,
$S_{cr}' < \left<S\right>$.  We have fixed the value of the exponent
$\kappa = 3$ in our simulations, which corresponds to solid spheres, according
to~\cite{delichatsios_model_1975}.

To facilitate our analysis we highlight here some of the relations between the
shear distribution of the random flow and the sizes of aggregates produced. The
PDF of shear, obtained from space sampling of $\mathbf{u} (\mathbf{X},t)$, is
shown in Fig.~\ref{fig:hist_shear} (a). For shear values below average it can be
described as a linearly increasing function:
PDF$(S < \left< S \right>) \propto S^{\chi}$, with $\chi = 1$.  Otherwise for
$S > \left< S \right>$ the probability distribution decreases sharply and has a
cutoff at $S_{\text{max}} \sim 3 \left<S\right>$, which is an approximation of
the asymptotic cutoff seen in Fig.~\ref{fig:hist_shear} (a). This aspect differs
significantly from the behaviour of a homogeneous and isotropic turbulent flow,
where the shear distribution assumes a broader shape, and has an exponential
decay for $S > \left< S \right>$~\cite{zeff_measuring_2003}, reaching values as
large as hundred times of the average shear. This can be noticed from the
appearance of extreme and rare events where the shear value suddenly rises. Such
events are specific to turbulence and are absent in random flows. For more
details about this comparison and how it reflects on fragmentation rates of
tracer particles, see~\cite{babler_numerical_2015}.

With a fixed $\kappa$ ($\kappa = 3$), we can attribute to each $\gamma$
corresponding size classes which facilitate the connection of the aggregate's
size distribution to the properties of the flow: $\alpha_{\left<S\right>}$ --
featured as a dashed blue line on the size distribution in
Fig.~\ref{fig:hist_shear} (b), corresponds to the size class with
$S_{\text{cr}} = \left<S\right>$; and $\alpha_{S_{\text{max}}}$ -- represented
as a dashed red line in Fig.~\ref{fig:hist_shear} (b), shows the size class with
$S_{\text{cr}} = S_{\text{max}}$. It is also important to note that the smallest
size in the suspension in steady state is $\sim \alpha_{S_{\text{max}}}/2$ since
we work with binary fragmentation where the resultant fragments are nearly of
the same size, except for small fluctuations.

Let us discuss the limiting case of the broadest size distribution of
aggregates. Since this size distribution depends only on the properties of the
flow, we take a snapshot of the flow field at one arbitrary time instant. For
this snapshot we determine the spatial distribution of the shear forces in the
flow. Suppose we distribute to each location in space one aggregate whose size
is the largest possible i.e. its size $\alpha$ is determined by the critical
shear at its position.  As an example, let the considered position be
$(x_0, y_0)$ and the shear at this position be
$S(x_0, y_0) = \left(2\sum_i \sum_j S_{ij}(x_0, y_0) S_{ij}(x_0,
  y_0)\right)^{1/2}$, then the corresponding size $\alpha$ at this position
would be $\alpha(x_0, y_0) =\left(\frac{1}{\gamma} S(x_0,
  y_0)\right)^{-3}$. Taking now aggregates on all positions $(x, y)$ in
configuration space fulfilling the condition of the maximum possible size before
breakup we obtain the broadest possible size distribution. In case of our random
flow for $\alpha > \alpha_{\left<S\right>}$ this strong aggregation limit,
taking into account that PDF$(S < \left< S \right>) \propto S^{\chi}$, produces
a size distribution $N_{\alpha} \propto \alpha^{-(\chi + \kappa +
  1)/\kappa}$. Which for $\kappa = 3$ and $\chi = 1$, has the form
$N_\alpha \propto \alpha^{-5/3}$ (see Fig.~\ref{fig:hist_shear} (b) and red
stars in Fig.~\ref{fig:SD_inertial} (a)).  This power law is the broadest size
distribution allowed by the shear of the flow. In the next section we show that
decreasing the dilution rate or increasing particle's inertia and particle's
segregation brings the size distribution closer to this limit.

\begin{figure}[h!]
  \includegraphics[scale=.64]{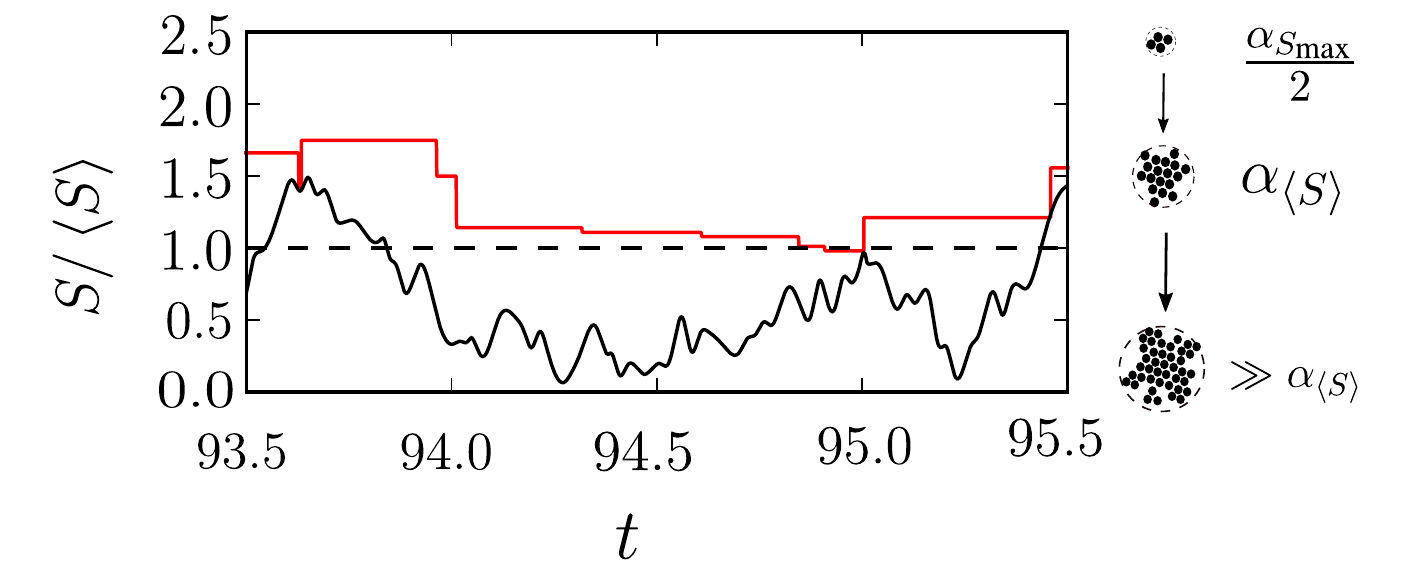}
  \caption{Schematic representation of the changes in critical shear attributed
    to an aggregate, which changes due to aggregation and fragmentation events
    (red), in comparison to the shear force of the flow at the aggregate's
    position (black).}\label{fig:follow}
\end{figure}

Finally, we would like to emphasize that the lifetime of an aggregate does not
only depend on its size, but also correlates with the time scales of the flow
field. For this purpose it is useful to follow an individual aggregate (in what
is also known as the Lagrangian view). We can monitor the shear forces this
aggregate experiences while advected by the flow field, and additionally, how
this aggregate changes its size during this time due to aggregation and
fragmentation events. Those events lead to transitions between size classes, see
Fig.~\ref{fig:follow}. This view allows us to compare the time scales between
aggregation-fragmentation events and the time scale of the flow. In fact, for a
random flow the Lagrangian velocity becomes uncorrelated after a time span
proportional to the correlation time $\tau_{f}$ or an eddy turnover time
$\lambda_f/u_0$. Therefore the sampling of the possible shear values depends
only on these two time scales in this simple set-up. For instance a tracer
starting at a region of the flow with low shear will, after a time interval
proportional to $\tau_{f}$, move into a region with high shear value. As a
consequence, the time an aggregate has to grow is finite, and restricted by the
flow time scales. Here we argue that the number of aggregation events happening
while the aggregate is entering a region of small shear is crucial to describe
the tail of the aggregate size distribution. Furthermore, the lifetimes of large
aggregates get smaller than $\tau_f$, beyond $\alpha_{\left<S\right>}$. In other
words such large particles are fragmented by the flow in a time span smaller
than the correlation time of the flow, Fig.~\ref{fig:follow}.

\section{Results}\label{sec:results}

\subsection{Dilution rate}\label{sec:dilution_rate}

\begin{figure}[t!]
  \includegraphics[scale=.65]{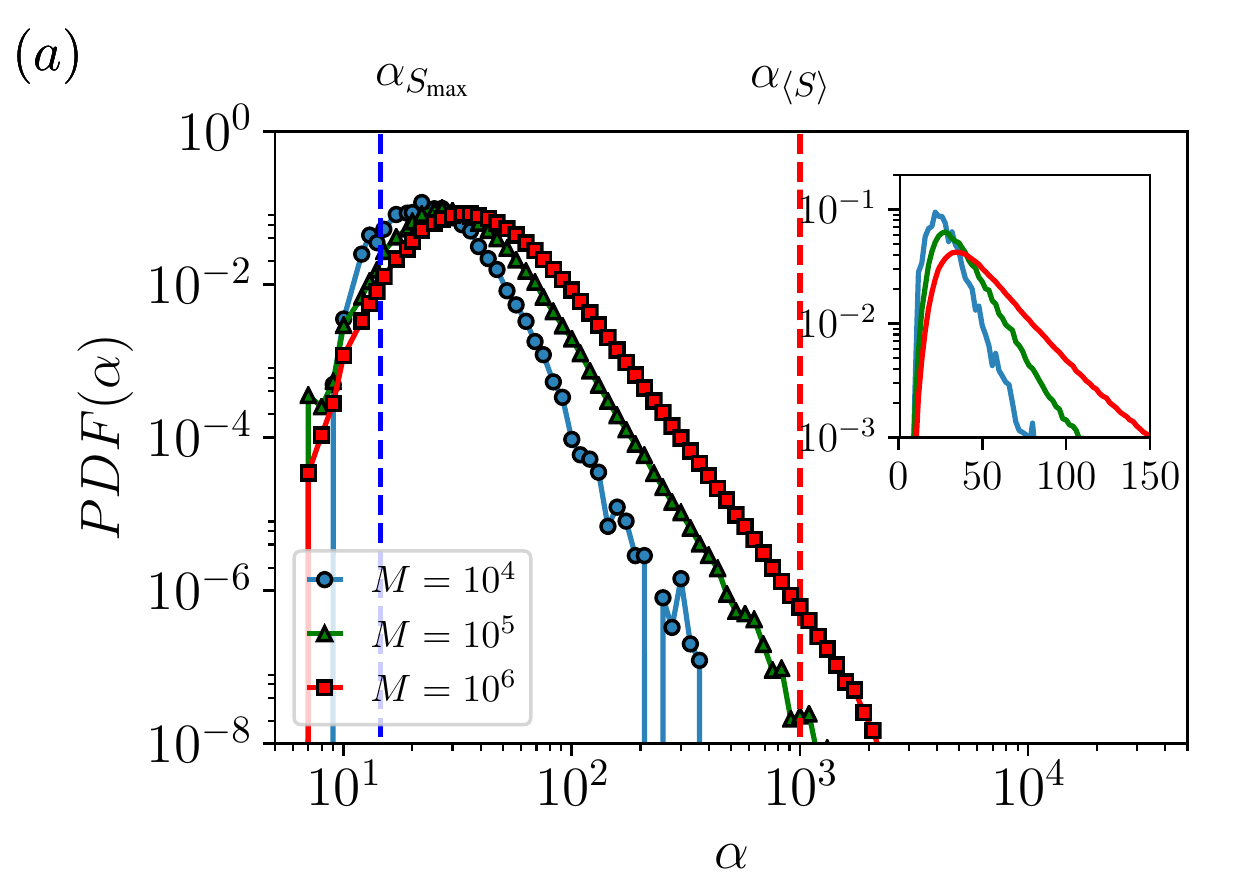}
  \includegraphics[scale=.65]{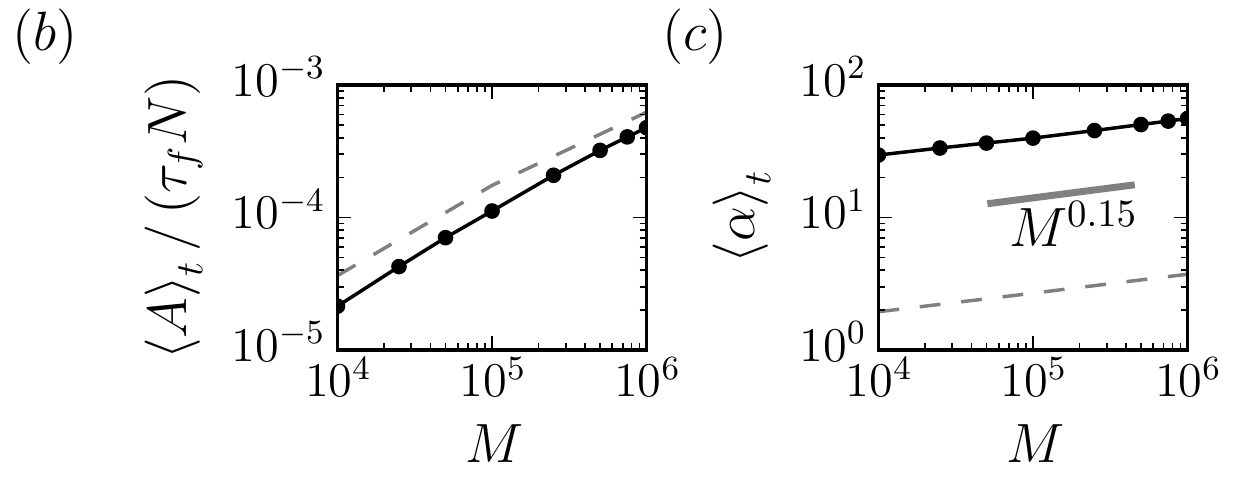}
  \caption{(a) Steady state size distribution of non inertial (tracer)
    aggregates obtained by changing the number of monomers in the flow, in the
    inset the same in log-lin representation; (b) Number of aggregation (or
    fragmentation) events per aggregate in one correlation time of the flow; (c)
    Average aggregate size as a function of the total number of monomers in the
    suspension. Tracer monomers characterized by a size
    $r_0 = 5 \cdot 10^{-4} \lambda_f$ and binding strength $\gamma = 10$ (solid
    lines) and $\gamma = 4$ (gray dashed lines in (b) and
    (c)).}\label{fig:SD_tracers}
\end{figure}

In this Subsection we compare the steady state size distributions obtained for
tracers and inertial monomers. We analyze the influence of the total number of
monomers $M$ on the aggregation and fragmentation processes, and the resulting
size distribution in the steady state.

We start by showing the resulting steady state size distributions for tracer
monomers, Fig.~\ref{fig:SD_tracers} (a). It is important to highlight two main
characteristics which distinguish the dynamics of tracers from inertial
aggregates: (1) the size of a tracer aggregate does not influence the advection
of this aggregate by the flow; (2) tracer aggregates are distributed uniformly
in space when advected by an incompressible flow field. Taking into account
these aspects, the appearance of spatial inhomogeneities in the concentration of
aggregates, for the tracer system can only result from aggregation and
fragmentation dynamics. In view of decreasing the simulation time, a large part
of these simulations for tracer monomers were run with $\gamma = 10$.  As
explained previously (see Sec.~\ref{sec:shear}) the size distribution is
characterized by a minimum size $\sim \alpha_{S_{\text{max}}/2}$, where
$S_{\text{max}}$ is a function of $\gamma$. As we increase the number of
monomers in the suspension, $\left<\alpha\right>_t$ increases and the size
distribution gets broader. We identify that the average size scales as
$\left<\alpha\right>_t \propto M^{0.15}$, Fig.~\ref{fig:SD_tracers}(c).  The
differences are not only observed in the averages, but also in the dynamics of
the steady state, which also changes with the number of monomers in the
suspension: we observe that the number of aggregation (fragmentation) events
increases with $M$. We have counted the total number of aggregation events in
the system in a correlation time $A$, and obtained an average for the number of
these events per aggregate in one correlation time in the steady state
$\left<A/N\right>_{t}$, where $N = \sum_{\alpha} N_{\alpha}$. In
Fig.~\ref{fig:SD_tracers}(b), we show how the average number of events per
aggregate changes with the increase of total number of monomers in the
system. Note that for the steady state the number of fragmentation events per
aggregate is on average the same as of aggregation events.

\begin{figure}[h!]
  \includegraphics[scale=.65]{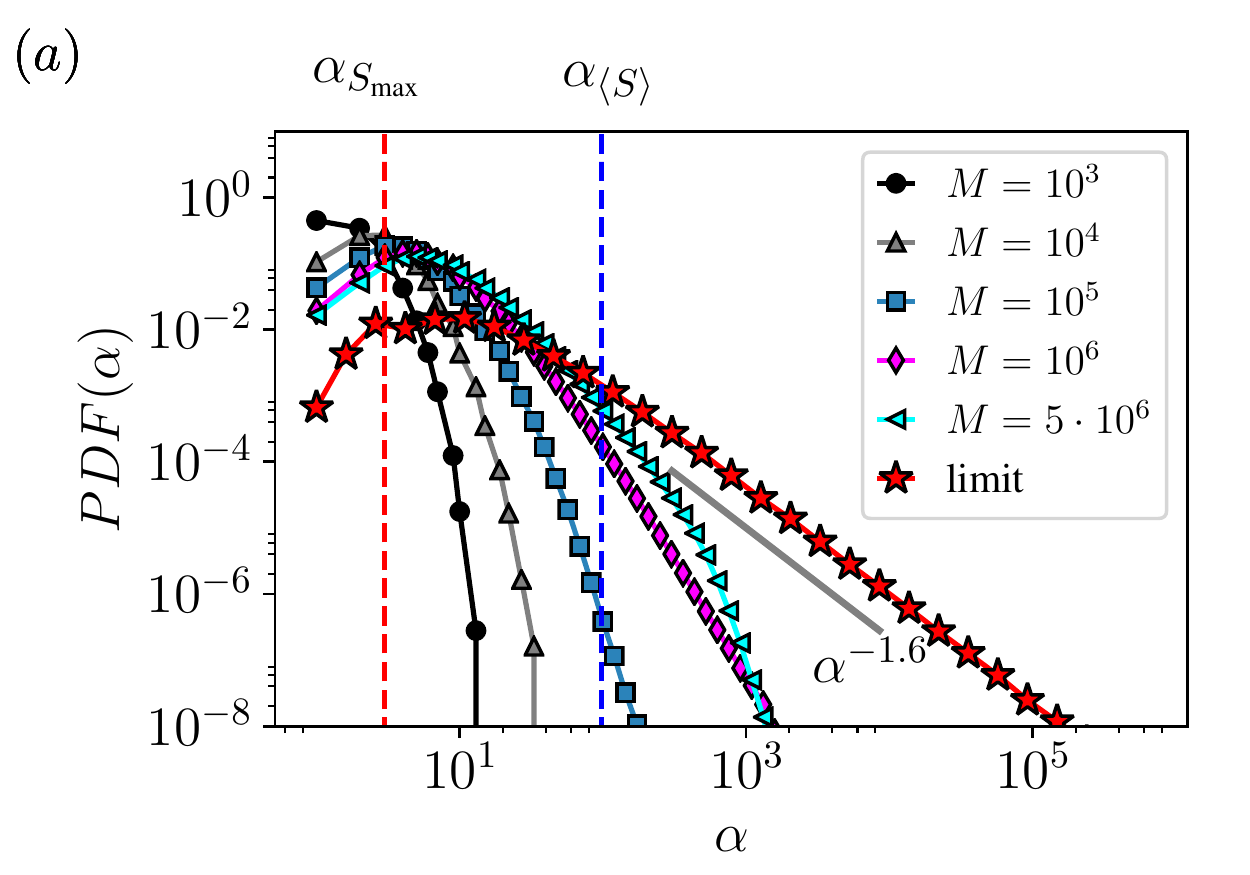}
  \includegraphics[scale=.65]{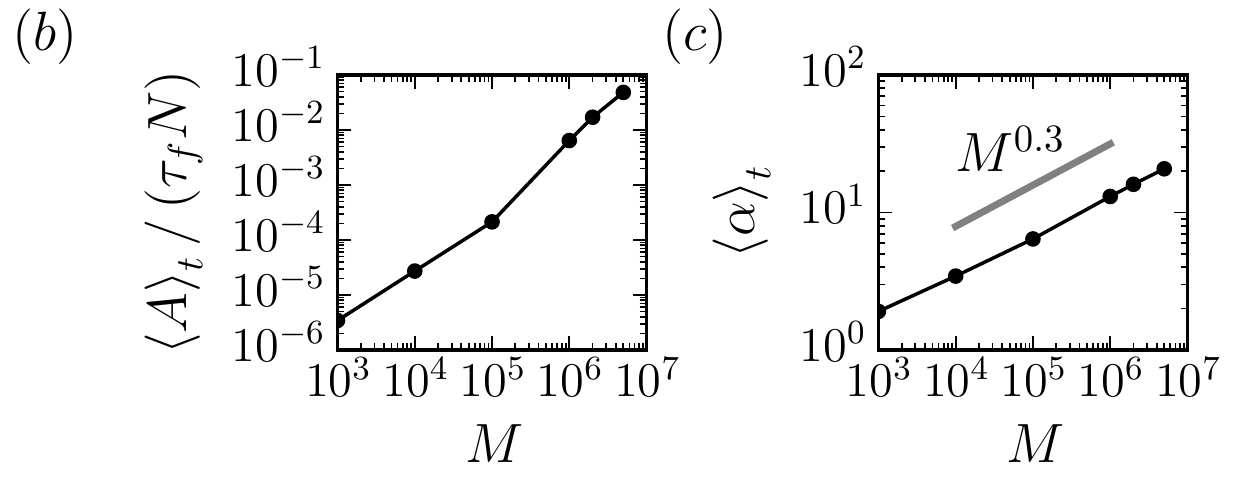}
  \caption{(a) Size distributions of aggregates initialized with different
    numbers of monomers $M$, for monomers with
    $r_p = 5 \cdot 10^{-4} \lambda_f$, $\beta = 0.1$, which corresponds to
    $St = 0.8$ and binding strength $\gamma = 4$. The size distribution
    represented with red stars corresponds to the limit distribution, described
    in Sec.II C. (b) Number of aggregation (or fragmentation) events per
    aggregate in one correlation time of the flow; (c) Average aggregate size as
    a function of the total number of monomers in the
    suspension.}\label{fig:SD_inertial}
\end{figure}

For a system with inertial monomers not only aggregation and fragmentation
dynamics depend on the position of these aggregates, but the advection dynamics
depends on the aggregate's size too. Furthermore, the spatial fluctuations in
aggregate numbers may be additionally enhanced by inertial effects, such as
preferential concentration, which in turn affects the collision rates. As in
previous paragraph we address the question of how the total number of monomers
influences the size distributions of aggregates.

The steady state that results from interactions of inertial aggregates in
strongly diluted suspensions with $M = 10^{3}$ monomers, is characterized by a
size distribution with a tail that appears to exhibit an exponential decay
(Fig.~\ref{fig:SD_inertial} (a)). The steady state here is characterized by a
small number of aggregation events $A$ per aggregate in a correlation time of
the flow, see Fig.~\ref{fig:SD_inertial} (b). As we increase the total number of
monomers the effects of inertia get more noticeable. These effects are reflected
in the average size of aggregates in the suspension, which for inertial
aggregates increases as $M^{0.3}$, see Fig.~\ref{fig:SD_inertial} (c).  The
observed broad size distributions are a result of: an enhancement of collisions
due to distinct reactions times that the aggregates of different sizes have to
the flow field and also due to preferential concentration, which is effective
only for some of the size classes.  As we have discussed in Sec.~\ref{sec:shear}
the fragmentation dynamics is coupled to the flow field, which has a preassigned
correlation time $\tau_f$. To grow up to sizes larger than
$\alpha_{\left<S\right>}$, aggregates have to experience several aggregation
events within $\tau_f$. Therefore a simple consequence of large numbers of
monomers (such as for $M = 10^{6}$, see Fig.~\ref{fig:SD_inertial} (a)) is a
broad distribution tail, with sizes that can reach values larger than
$\alpha_{\left<S\right>}$. However, from changes in the shape of the size
distribution we can notice that the relation between aggregation and
fragmentation dynamics is non trivial. Although, the number of both type of
events are the same on average, the sizes involved in aggregation and
fragmentation dynamics change. We can observe that large aggregates need longer
times to form, since they result from sequences of aggregation events. However,
they are characterized by very short ``lifetimes'' (much shorter than
$\tau_f$). As a consequence, even if large aggregates are formed in a
correlation time they are quickly fragmented after their formation. In spite
their short appearance in the suspension, their formation transforms the size
distribution's tail into a power-law-like shape.

As a result, for the case of tracers and especially of inertial aggregates, the
shape of the size distribution depends strongly on the dilution rate of the
suspension.

\subsection{Influence of the binding strength}\label{sec:binding}
\subsubsection{Tracer monomers}\label{sec:binding_tracers}

As stated previously the main parameter that linearly changes the resistance of
all aggregates to shear forces in the flow, is the binding strength among
monomers, represented by parameter $\gamma$.  While an increase in $\gamma$ also
means that aggregates are more resistant to fragmentation, the result is a size
distribution with a larger mean number of aggregates, $\left<\alpha \right>_t$,
in the steady state of the process. Furthermore, as $\gamma$ increases we create
aggregates with $\alpha > 1$, which are resistant even to the strongest
hydrodynamic forces exerted by the flow field. These aggregates end up
functioning as the smallest units in the ensemble, and can therefore be
considered as ``effective monomers''. For an ensemble of tracers these
``effective monomers'' will have all properties identical to the original
monomers, with the exception to their radius, $r_{\alpha} > r_p$. The outcome is
a similar system, but with a smaller number of ``effective monomers'' and a
different relation between ``effective monomer'' size $r_{\alpha}$ and the size
of the vortices $\lambda_f$.

\begin{figure}[h!]
  \includegraphics[scale=.7]{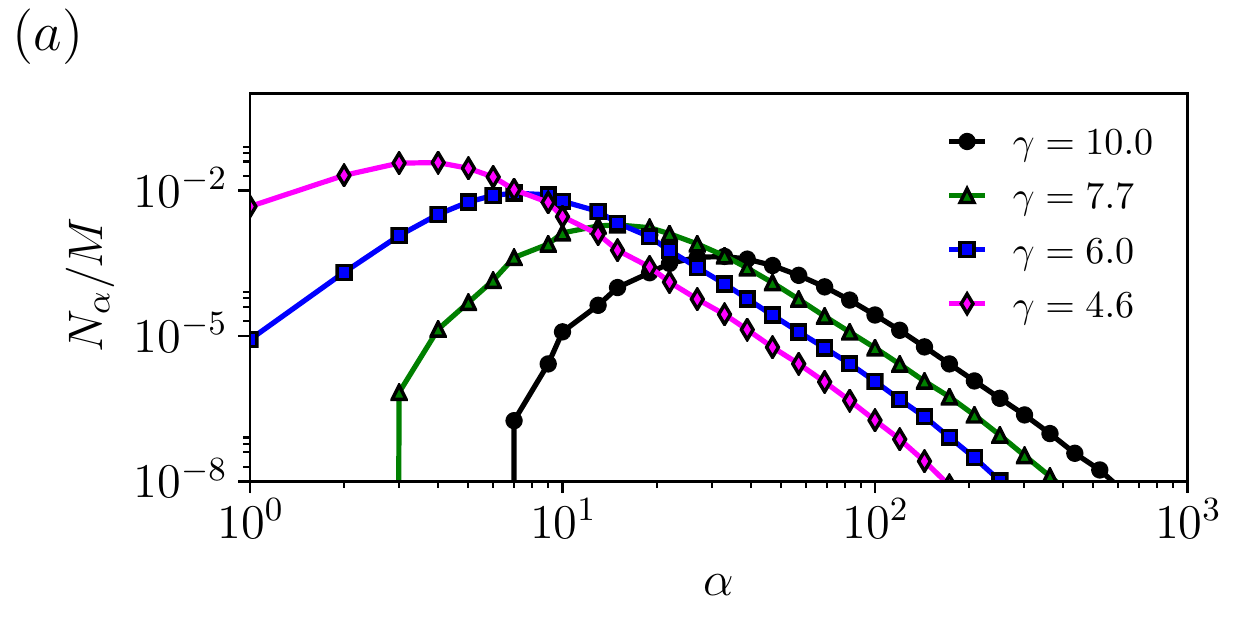}
    \includegraphics[scale=.7]{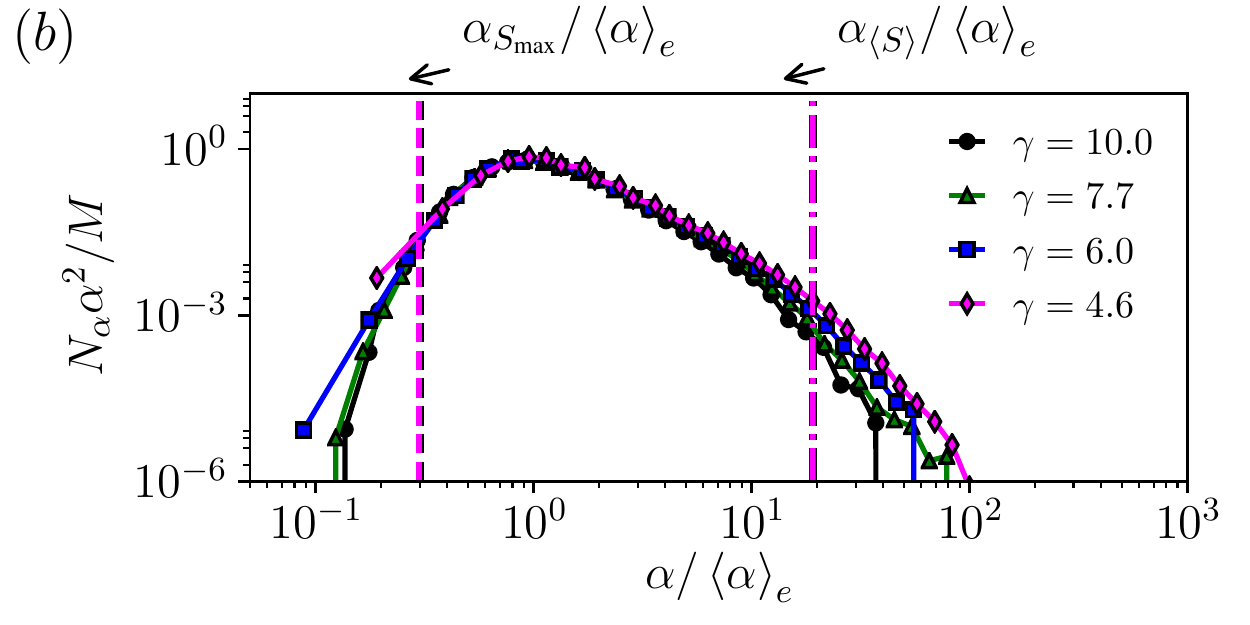}
    \caption{Steady state size distributions of tracer aggregates for different
      binding strengths $\gamma$. Tracer monomers are characterized by a size
      $r_p = 5 \cdot 10^{-4} \lambda_f$ in a suspension with $M = 10^6$.  (a)
      The size distribution in its original form. (b) Rescaled size distribution
      according to Eq.(\ref{eq:scaling}).  }\label{fig:SD_tracers_gamma}
\end{figure}

For tracers we chose to work with $M = 10^6$, since their dynamics is
characterized by long transients. The large duration of transients is a
consequence of small collision rates, due to the homogeneous spread of
aggregates in space. Here we use the relation given by Eq.(\ref{eq:scaling}), to
rescale the obtained size distributions. The size distribution in this form
possesses a horizontal axis which is divided by $\left<\alpha\right>_e$, a
quantity that also grows with $\gamma$. The scaling of $\left<\alpha\right>_e$
with the binding strength is discussed in more detail in Subsec. III B3. The
tendency for all our study cases, therefore, is to have broader distributions
for larger $\gamma$ values (Fig.~\ref{fig:SD_tracers_gamma} (a)). In particular
for tracers the average and the standard deviation (first and second moments of
the size distribution) grow with the same proportion. We obtain a partial
overlap of distributions in their rescaled form for different values of
$\gamma$, Fig.~\ref{fig:SD_tracers_gamma} (b), with deviations only in the tail
of the distribution. In this case, the size distribution has a sharper decrease
towards larger aggregates and fewer aggregates that grow beyond
$\alpha_{\left<S\right>}$.  The maximum value is only four times larger than
$\alpha_{S_{\text{max}}}$ and we observe a narrow distribution. Note that for
all dilution rates used, $\left<\alpha\right>_e$ is only slightly above
$\sim \alpha_{S_{\text{max}}/2}$.  Furthermore, the largest fraction of the
system's monomers remains in aggregates of sizes below
$\sim \alpha_{\left<S\right>}$, characterizing a regime of few aggregation
events per $\tau_f$.  The steady state size distribution can be rescaled with
respect to the characteristic size of aggregates in the steady state, which in
this case can be either the average aggregate $\left<\alpha\right>_e$ or the
``effective monomer''.

\subsubsection{Inertial monomers}\label{sec:binding_inertial}

\begin{figure}[h!]
  \includegraphics[scale=.7]{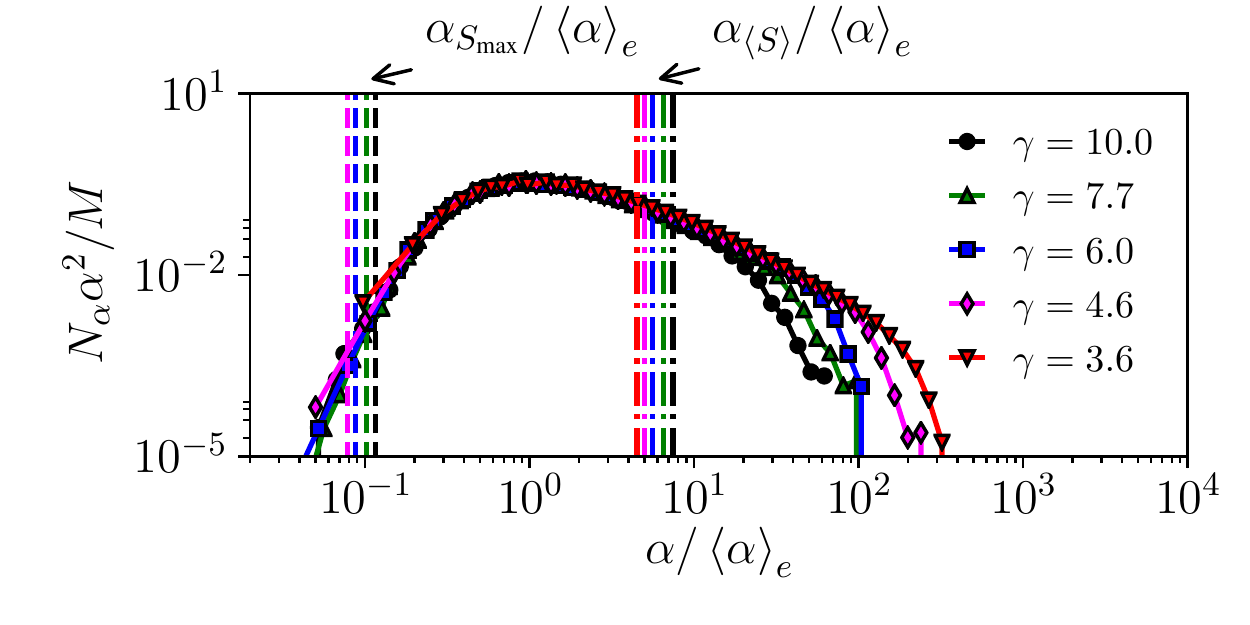}  
  \caption{Steady state size distributions $N_{\alpha}$ changing $\gamma$,
    binding strength among monomers. The data is scaled using the average
    aggregate size in the distribution $\left<\alpha\right>_e$, according to
    Eq.(\ref{eq:scaling}.  The simulations contains inertial aggregates
    ($St = 0.83$, $\beta = 0.1$) in suspension with $M = 10^{6}$
    monomers.}\label{fig:size_distrib_inertial_gamma}
\end{figure}

Next we compare the steady state distribution for ensembles of inertial monomers
characterized by different binding strengths. We initialize a suspension with
$M = 10^{6}$ monomers.  We again check the rescaled form of the size
distribution, according to Eq.(\ref{eq:scaling}), using $\left<\alpha\right>_e$
as the characteristic size. As for the case of tracers the use of this scaling
collapses all size distributions with deviations appearing only in the tail, see
Fig.~\ref{fig:size_distrib_inertial_gamma}.

The aggregation and fragmentation dynamics produces a mass flux flowing towards
sizes much larger than $\alpha_{\left<S\right>}$.  As we described in
Sec.~\ref{sec:shear}, for this set-up an aggregate may have sufficient time to
grow before the flow decorrelates and exposes it to a high shear rate. Here one
correlation time of the flow is enough for aggregates to grow into sizes much
larger than $\alpha_{\left<S\right>}$. We observe that the flux of mass towards
sizes larger than $\alpha_{\left<S\right>}$ leads to size distributions with a
significantly broader tail for small $\gamma$, after the rescaling of the
distribution. It is important to note that the lifetime of an aggregate of size
larger than $\alpha_{\left<S\right>}$ is shorter than $\tau_f$, and therefore
all large aggregates, which form the tail of the size distribution, have a short
free path before a breakage event occurs. Nevertheless the formation of these
aggregates changes strongly the shape of the size distribution.

\subsubsection{From tracers to inertial aggregates: intermediate cases.}\label{sec:tr_vs_inertial}
In the previous subsections we have analyzed two limiting cases: aggregation of
tracers and aggregation of inertial monomers. In these two setups we have
observed that obtained aggregate size distributions in the steady state for
different $\gamma$ values can all be rescaled by a characteristic aggregate
size, $\left<\alpha\right>_e$ suggested from Eq.(\ref{eq:scaling}). This average
aggregate size $\left<\alpha\right>_e$, in turn grows algebraically with the
binding strength. Here in this final subsection we study the scaling properties
of the steady state size distribution for suspensions of aggregates with
intermediate inertial properties and show cases where the size distribution
changes its shape with changes in $\gamma$. We show simple examples where the
dynamics of an average size is equivalent to tracers for small $\gamma$ values
and to inertial aggregate for large $\gamma$ values. We speculate that the
enhancement of aggregation rates for inertial monomers arises from the interplay
of two effects: it is, more importantly, a result of differences in reaction
times in the advection between aggregates of different sizes (as the range of
sizes increases so does the range of Stokes times $\tau_{\alpha}$); and also, to
a smaller degree, a result of preferential concentration, which is significant
for some size classes only. Although the second phenomenon has a strong impact
only for aggregates with Stokes time $\sim 1$, the first one depends on the
variety of sizes in the suspension.  Here we analyze in detail this interplay,
and to this end we introduce two additional monomer types, which exhibit
intermediate properties between tracers and inertial aggregates: small inertial
monomers ($St = 0.08$ and $\beta = 0.1$); and weakly inertial monomers
($St = 0.08$ and $\beta = 0.99$).

\begin{figure}[h!]
  \includegraphics[scale=1.]{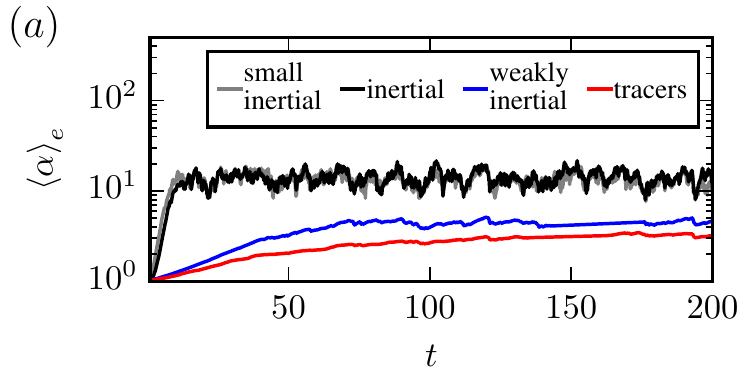}
  \includegraphics[scale=1.]{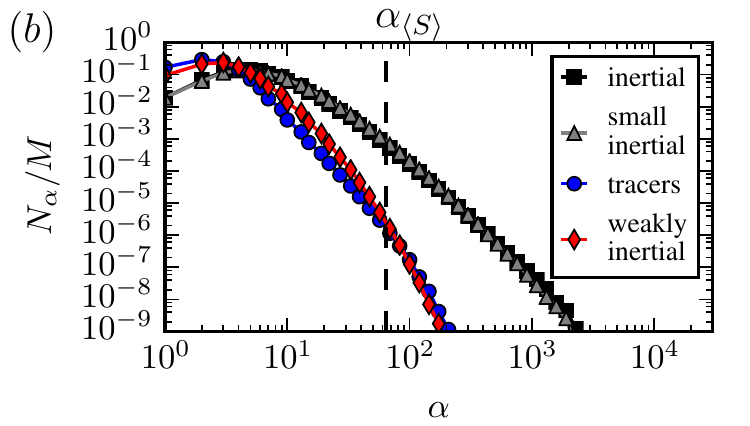}
  \caption{(a) Time evolution of the average aggregate size in the
    suspension. (b) Size distribution of aggregates in the steady
    state. Each ensemble contained $10^{6}$ monomers and the binding
    strength between monomers was $\gamma = 4$.}\label{fig:agg_beta}
\end{figure}

Firstly, we compare the aggregation dynamics of tracers and for inertial
particles, and distinguish the four types of monomers: (i) inertial, (ii) small
inertial, (iii) weakly inertial and (iv) tracers. The differences can be
observed already in the transient dynamics. As we can see from
Fig.~\ref{fig:agg_beta} (a), the transients in the dynamics of small inertial
and inertial monomers (cases (i) and (ii), almost indistinguishable in
Fig.~\ref{fig:agg_beta} (a)) are much shorter than of the weakly inertial and
tracer cases.  Inertial aggregates also grow to larger sizes, and the mean size
has stronger and more frequent fluctuations, indicating the presence of a larger
number of aggregation and fragmentation events. The short transients and a large
average size are both consequences of an increase in collision rates due to
inertial effects. The mass quickly flows towards larger sizes, and we obtain a
broad size distribution, see squares in Fig.~\ref{fig:agg_beta} (b).

\begin{figure}[h!]
  \includegraphics[scale=.7]{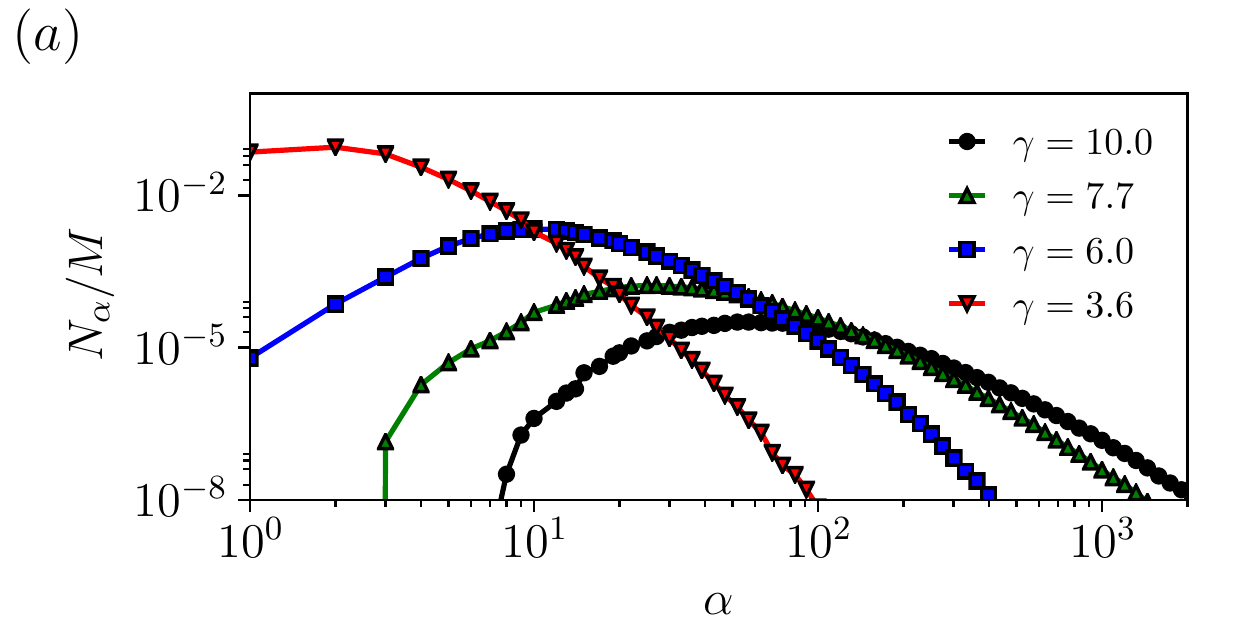}
  \includegraphics[scale=.7]{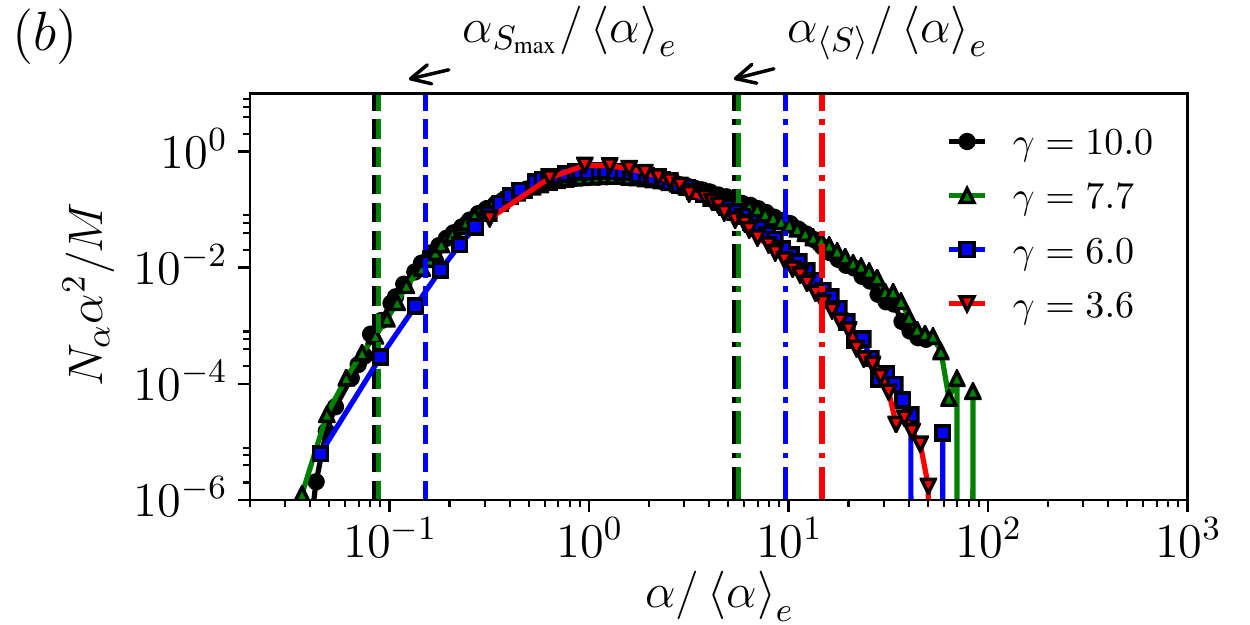}
  \caption{Steady state size distribution for weakly inertial monomers
    ($St = 0.08$, $\beta = 0.99$). The simulation contains $10^{6}$ monomers.
    (a) The size distribution in its original form. (b) Rescaled size
    distribution according to Eq.(\ref{eq:scaling}).}\label{fig:size_distrib_w}
\end{figure}

The average aggregate size $\left<\alpha\right>_e$ in the suspension is
comparatively smaller for both tracer monomers (black line in
Fig.~\ref{fig:agg_beta} (a)) and weakly inertial monomers (blue line in
Fig.~\ref{fig:agg_beta} (a)). This is because, as tracers, weakly inertial
monomers also have the tendency to distribute themselves nearly homogeneously in
space, while advected by an incompressible flow. Although, for the chosen value
of $\gamma$, inertial effects appear to be small, we can still notice their
influence in both the duration of transients and the average aggregate size, see
Fig.~\ref{fig:agg_beta}.  While tracers collide only due to spatial variations
of the velocity field ~\cite{saffman_collision_1956}, the drag force makes
trajectories of aggregates formed by weakly inertial monomers deviate from the
underlying flow. As differences in sizes of these aggregates increase, so does
the range of Stokes times of these aggregates. They all react differently to the
velocity field, which results in an enhancement of collision rates. This
influences the average size and the shape of the aggregate size distribution in
the steady state, Fig.~\ref{fig:agg_beta} and Fig.~\ref{fig:size_distrib_w}. It
is important to emphasize that the inertial effects grow with an increase in the
variety of different sizes in the ensemble.  This in turn boosts the flux of
mass towards large sizes generating a positive feedback.

\begin{figure}[h!]
   \includegraphics[scale=.8]{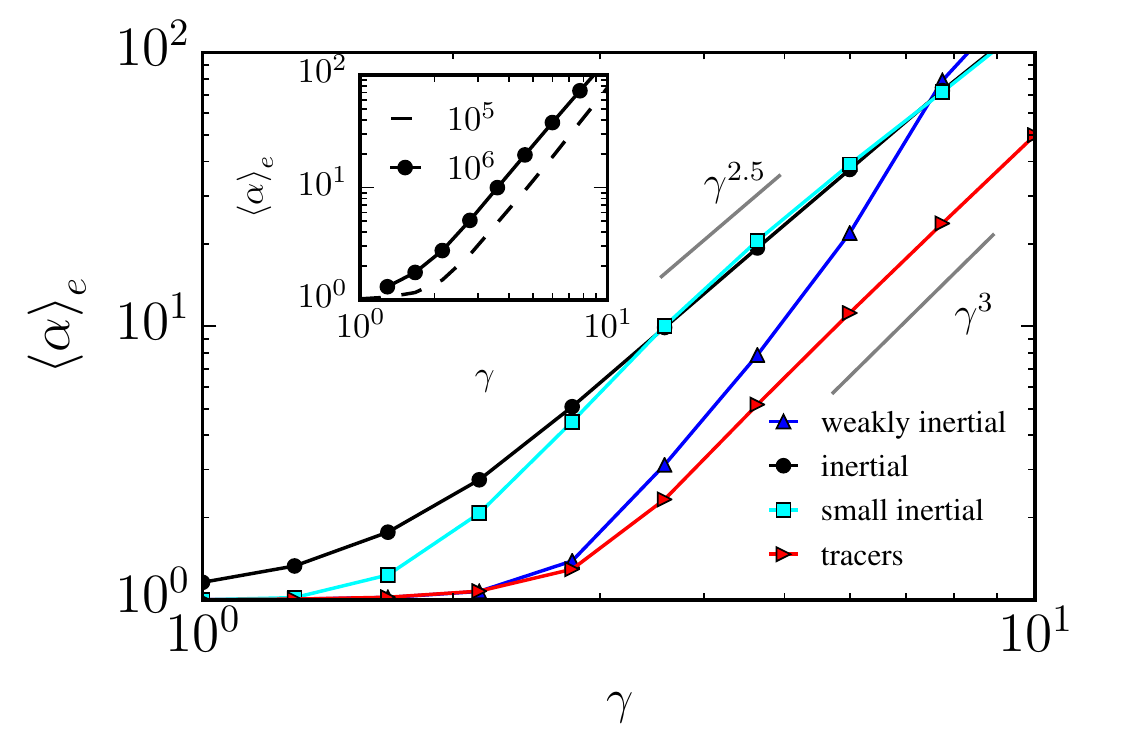}\\
   \caption{Average aggregate size in the steady state $\left<\alpha\right>_t$
     for inertial ($St = 0.83$, $\beta = 0.1$), small inertial ($St = 0.08$,
     $\beta = 0.1$), weakly inertial ($St = 0.08$, $\beta = 0.99$) and tracer
     aggregates changing the binding strength $\gamma$, the simulation was
     initialized with $M = 10^{6}$. The inset shows $\left<\alpha\right>_t$ for
     a system with $10^{6}$ (black dots) and another with $10^{5}$ (dashed line)
     monomers, characterized by $\beta = 0.1$.}\label{fig:gamma_summary}
\end{figure}

Secondly we compare the effect of variations of binding strength for tracers and
our three types of inertial aggregates. These results are summarized in
Fig.~\ref{fig:gamma_summary}. The average aggregate size
$\left<\alpha \right>_t$ always grows with $\gamma$, but as already demonstrated
for the size distribution the steady state may show different properties for
small and large values of $\gamma$. Fig.~\ref{fig:gamma_summary} shows that the
mean aggregate size $\left<\alpha\right>_t$ scales as $ \gamma^{\omega}$ (with
$\omega \sim 2.5$ for inertial aggregates and $\omega \sim 3$ for tracers). The
scaling of the average size with $\gamma$ for the suspension with tracer
monomers, directly reflects the fact that the smallest aggregate in the
suspension grows as $\propto \gamma^{3}$ and the dynamics of motion for these
aggregates does not depend on their size. By contrast, for inertial monomers the
properties of smallest aggregates in the suspension do change. As the
``effective monomers'', the smallest unbreakable aggregates, increase in size
with $\gamma$ they start to spread homogeneously in space, and preferential
  concentration is decreased. However, the differences in Stokes times of
aggregates formed in the suspension still have a strong effect in keeping the
aggregation rates higher than that of these tracer monomers. A change in
dynamics is also noticeable for small inertial monomers (with $St = 0.08$,
$\beta = 0.1$). For small $\gamma$ they behave as tracers but with stronger
binding forces they can form ``effective monomers'' that are large enough to
experience the influence of the drag. At this value of binding strength the
average aggregate produced in the steady state by small inertial monomers is
identical to the one produced by aggregation-fragmentation dynamics of inertial
monomers ($St = 0.83$, $\beta = 0.1$). For weakly inertial monomers (blue
triangles Fig.~\ref{fig:gamma_summary}) this transition in the dynamics is even
more pronounced. Here the increase of binding strength leads to a strong
modification of the shape of the steady state size distributions. As a
consequence, size distributions formed by monomers possessing different values
of $\gamma$ cannot be rescaled into a similar form with the use of
Eq.(\ref{eq:scaling}), see Fig.~\ref{fig:size_distrib_w}.

\section{Conclusion}~\label{sec:conclusion}

We have investigated aggregation and fragmentation dynamics of tracers and
inertial aggregates in random flows. We have used an individual particle based
model to compute the motion of aggregates, characterizing each one of them by
its coordinates in space, its velocity and its size. Furthermore, the
computation of individual trajectories made it possible to exactly evaluate
breakup events that depend on local hydrodynamic forces. Throughout this work we
have compared properties of the steady-state for several ensembles, where we
have varied the monomer's Stokes number and density. Our objective was to
analyze the influence of aggregation on the steady-state size distributions,
which result from differences in the advection dynamics. While the equations of
motion are independent of aggregate sizes for tracer aggregates, for inertial
ones they depend on the sizes present in the suspension. We also analyzed the
dependency on the number of monomers (i.e. the solid fraction) and their binding
strength, both of which change the average aggregate size in the steady-state
$\left<\alpha\right>_t$, albeit in different ways.

We observe that beside the expected result that the size distribution gets
broader for all ensemble types as the number of monomers increases, we also
observe that the shape of these size distributions change as well. These changes
are especially noticeable in the distribution tail, while the average aggregate
size grows with $M^{0.15}$ for tracers and $M^{0.3}$ for inertial aggregates. In
contrast to predictions by mean field theory we do not find a characteristic
size able to rescale the distribution into a single shape. We interpret that
these differences in the shape of the distribution follow from an interplay
between the collision and flow time scales, since with an increase in $M$ more
aggregation events can occur within the correlation time of the flow. In
addition to this, we have also the correlation between particle sizes and the
local magnitude of the shear forces, which should also contribute to the
deviation from what is expected in a mean-field scenario.

The increase in the binding strength also results in broader size distributions.
Additionally, we have observed that for each one of the two limiting cases ---
tracer and inertial monomers --- it is possible to identify a characteristic
size which allows us to rescale the size distribution into an almost universal
shape (with deviations only in the tail). In case of tracers, this
characteristic size is either the smallest particle in the steady state (the
``effective monomer''), or the average aggregate size in the steady state, which
grows as $ \gamma^{3}$. In the case of inertial monomers, only the average
aggregate size can be used as the characteristic size for the scaling, which in
turn grows as $\gamma^{2.5}$. This smaller exponent is a result of a decrease in
the preferential concentration experienced by the ``effective monomers''. The
weaker preferential concentration for larger $\gamma$ values brings the
average size closer to the one characterizing a system of tracers, where the
spatial distribution of aggregates is homogeneous. Finally, we have also
analyzed monomers with intermediate inertial properties, i.e. small Stokes
numbers and almost neutrally buoyant. In both cases the scaling failed, since
with an increase in $\gamma$ these systems shift from a tracer-like to an
inertial-like behaviour. During this transition, the size distributions exhibits
different shapes, according to the inertial properties of the aggregates present
in the steady state.

Another aspect which we would like to bring to attention is that in real world
systems not all collision events result in aggregation, i.e. aggregation occurs
only with a certain probability corresponding to a collision
efficiency. Furthermore, there should be a direct relation between binding
strength $\gamma$ and the value for this collision efficiency, since both
represent interactions among parts of an aggregate. It was shown
in~\cite{zahnow_what_2009} that the effects of collision efficiency on the size
distribution is equivalent to the influence of the dilution rate.  Although in
this work we have decided for decoupling the two aspects, one can guess the
combined results from simultaneous changes in $\gamma$ and $M$.

In summary, we have showed how differences in the advection and subsequently in
collision dynamics impact on the steady state size distribution of aggregation
and fragmentation processes. We have demonstrated the applicability of the
scaling relation provided by the mean-filed theory of reversible aggregation for
our individual based approach and also demonstrated when this relation
fails. Our results show that it is important to take into account that inertial
aggregates change their advection dynamics with changes in their size.

Finally, we would like to briefly mention the possibility of applying this
approach to the formation and settling of marine snow. These aggregates are
almost neutrally buoyant, with densities close to the water density (as in our
case of weakly inertial aggregates). For this reason, these aggregates have an
almost homogeneous distribution in the flow~\cite{guseva_history_2016}. Through
a sequence of aggregation and fragmentation events, these aggregates reach a
given size distribution while advected by the flow field. Because they are very
light particles, the fragmentation events are triggered by the hydrodynamic
stress.  Although their horizontal motion and dispersion are similar to tracers,
they also sediment, and as a consequence their settling speed is determined by
the aggregate size, and thus is tightly linked to the aggregation and
fragmentation processes. It is therefore very important to correctly determine
the balance between aggregation and fragmentation to evaluate the settling
velocity for marine snow. Our results suggest that even for such light particles
the inertial properties of a characteristic aggregate size (average size) may
influence the overall shape of the size distribution.

\section{Acknowledgments}
We are grateful to George Jackson, Tamás Tél, Jöran März and Jens Zahnow for
illuminating discussions.

\appendix

\section{}

Here we describe the core aspects of the random velocity field (also know as
synthetic/kinematic turbulence) which was used for our simulations. It is based
on statistical properties of homogeneous, isotropic and stationary turbulence
and is constructed in such a way to contain predominately one length for its
coherent structures ($\lambda_f$) and be characterized by a single correlation
time scale ($\tau_f$). Therefore, although the random flow is unable to take
into account the interplay of the large variety of scales present in a turbulent
flow field, it mimics its main statistical properties such as correlations and
the energy spectrum with a lower computational effort.  More details of the
implementation and properties of random flows can be found
in~\cite{careta_stochastic_1993, marti_langevin_1997, sigurgeirsson_model_2002}.

For our simulations we assume a two dimensional (2D) and incompressible velocity
field ${\bf u}({\bf r}, t)$, which extends over a periodic box with dimensions
of $L\times L$. The incompressibility (${\bf \nabla} \cdot {\bf u} = 0$ ) of the
flow allows us to introduce a stream function ${\bf \psi}$ defined as:
\begin{equation}
  {\bf u}({\bf r}, t) = \left(\frac{\partial \psi}{\partial y}, -\frac{\partial \psi}{\partial x} \right).
\end{equation}
This stream function is assumed to have Gaussian, Markovian statistics and to be
describable by an Orstein-Uhlenbeck process. To establish proper spatial
correlations it is convenient to treat the process in Fourier space
\begin{equation}
  \psi({\bf r}, t) = \frac{\sqrt{\pi}}{L} \sum_{\bf k} \psi_{\bf k}(t) e^{i {\bf k}\cdot {\bf r}},
\end{equation}
\begin{equation}
  \psi_{\bf k}(t + \delta t) = \psi_{\bf k}(t)\left(1 - \frac{dt}{\tau_f}\right) + u_0 \lambda_f^2 Q(\lambda_f, {\bf k}) \sqrt{\frac{dt}{\tau_f}} w_{\bf k}.
\end{equation}
Where $\tau_f$ and $\lambda_f$ are respectively the correlation time and length,
$u_0$ is the characteristic velocity and $Q(\lambda_f, {\bf k})$ is chosen
according to the energy spectrum used. The noise is introduced through
$w_{\bf k}$, which are complex Gaussian random numbers which have zero mean and
are delta correlated,
\begin{equation}
\left<w_{\bf k}\right> = 0, \qquad w_{\bf k}* =  - w_{\bf k} \qquad \text{and} \qquad \left<w_{\bf k} w_{\bf k'}\right> = \delta_{\bf k k'}
\end{equation}
for more details see~\cite{garcia-ojalvo_generation_1992}.

The energy spectrum we consider in our simulation is the Kraichnan's spectrum:
\begin{equation}
  E(k) \propto  k^3  e^{- \lambda_f^2 k^2} = k^3 Q^2(\lambda_f, k).
\end{equation}
However, it is important to note that the spectrum is not well resolved for
$\frac{L}{\lambda_f} < 10$, and one single length scale dominates as the
characteristic vortex size.\\

\bibliographystyle{apsrev4-1}
\bibliography{aerosol}

\end{document}